\documentclass{ws-procs9x6}

\usepackage{amsmath}

\newcommand{\Tr}{\,\mathrm{Tr}\,}
\newcommand{\mev}{\,\mathrm{MeV}}

\begin{document}

\title{Progress in resolving charge symmetry violation in nucleon structure}

\author{R.~D.~Young$^*$, P.~E.~Shanahan and A.~W.~Thomas}

\address{ARC Centre of Excellence in Particle Physics at the Terascale and CSSM,\\
School of Chemistry and Physics, University of Adelaide, \\
Adelaide SA 5005, Australia\\
$^*$E-mail: ross.young@adelaide.edu.au}

\begin{abstract}
  Recent work unambiguously resolves the level of charge symmetry
  violation in moments of parton distributions using $2+1-$flavor
  lattice QCD. We introduce the methods used for that analysis by
  applying them to determine the strong contribution to the
  proton-neutron mass difference. We also summarize related work which
  reveals that the fraction of baryon spin which is carried by the
  quarks is in fact structure-dependent rather than universal across
  the baryon octet.
\end{abstract}


\bodymatter

\section{Introduction}

Charge symmetry, the equivalence of the $u$ quark in the proton and
the $d$ in the neutron, and vice versa, is an excellent approximation
in nuclear and hadronic systems --- typically respected at $\sim 1\%$
precision~\cite{Londergan:2009kj,Londergan:1998ai,Miller:2006tv}. Current deep inelastic
scattering measurements are such that this level of precision has not yet been
reached, with current bounds on charge symmetry violation (CSV) in
parton distributions in the range
5-10\%~\cite{Martin:2003sk}.
Such possibly large CSV effects are of particular interest in the
context of a new program at Jefferson Laboratory~\cite{JLAB} which
aims to measure the electron-deuteron parity-violating deep inelastic
scattering (PVDIS) asymmetry to better than 1\% precision. This would
offer an improvement of roughly an order of magnitude over early SLAC
measurements\cite{Prescott:1979dh}, with the potential to constitute
an important new test of the Standard Model. Reaching this goal will
rely on a precise control of strong interaction processes. CSV is
likely to be the most significant hadronic uncertainty at the kinematics typical
of the JLab
program\cite{Hobbs:2008mm,Hobbs:2011dy,Mantry:2010ki}.
Phenomenological studies suggest that CSV could cause $\sim 1.5-2\%$
variations in the PVDIS asymmetry~\cite{Martin:2003sk}. This is
sufficient to disguise any signature of new physics, such as
supersymmetry, expected to appear at the $1\%$
level~\cite{Kurylov:2003xa}.

Here we review our recent work\cite{Shanahan:2013vla} which has
determined the CSV moments of parton distributions from lattice QCD.
Our results, based on $2+1$-flavor lattice QCD
simulations~\cite{Horsley:2010th,Cloet:2012db}, reveal $\sim 0.20 \pm
0.06\%$ CSV in the quark momentum fractions.
This corresponds to a $\sim 0.4-0.6\%$ correction to the PVDIS asymmetry.
This precision represents an order of magnitude improvement over the
phenomenological bounds reported in Ref.~\cite{Martin:2003sk}.
This result also constitutes an important step towards resolving the
famous NuTeV anomaly~\cite{Zeller:2001hh,Bentz:2009yy}. Whereas the
original report of a 3-sigma discrepancy with the Standard Model was
based on the assumption of negligible CSV, effects of the magnitude
and sign reported here act to reduce this discrepancy by one sigma.
Similar results for spin-dependent parton CSV suggest corrections to
the Bjorken sum rule\cite{Alekseev:2010hc} at the half-percent level
which could possibly be seen at a future electron collider\cite{Deshpande:2005wd}.

In Section~\ref{sec:BaryonSplittings} we introduce the techniques used
for our calculation in the context of the octet baryon mass
splittings\cite{Shanahan:2013apa}.
Section~\ref{sec:CSVParton} summarizes our parton CSV results,
presented in full in Ref.~\cite{Shanahan:2013vla}.
Related work which reveals that the fraction of baryon spin which is
carried by the quarks is in fact structure-dependent rather than
universal across the baryon octet\cite{Shanahan:2012wa} is highlighted
in Section~\ref{sec:SpinFrac}.

\section{Baryon mass splittings}
\label{sec:BaryonSplittings}

Charge symmetry refers to the invariance of the strong interaction
under a $180^\circ$ rotation about the `2' axis in isospin space. At
the parton level this invariance implies the equivalence of the $u$
quark in the proton and the $d$ quark in the neutron, and
vice-versa. The symmetry would be exact if
\begin{itemize}
\item the up and down quarks were mass degenerate: $m_u=m_d$
\item the quark electromagnetic charges were equal: $Q_u=Q_d$.
\end{itemize}
Of course, both of these conditions are broken in nature. This
breaking manifests itself, for example, as mass differences between
members of baryon isospin multiplets.
While these differences have been measured extremely precisely
experimentally\cite{Beringer:1900zz},
the decomposition of these quantities into strong (from $m_u \ne m_d$)
and electromagnetic (EM) contributions is much less well known.
Phenomenological best estimates come from an application of the
Cottingham sum rule\cite{Gasser:1982ap} which relates the
electromagnetic baryon self-energy to electron scattering observables.
Walker-Loud, Carlson \& Miller (WLCM) have recently revised the
standard Cottingham formula\cite{WalkerLoud:2012bg}; noting that two
Lorentz equivalent decompositions of the $\gamma N \rightarrow \gamma
N$ Compton amplitude produce inequivalent self-energies,
WLCM use a subtracted dispersion relation to remove the
ambiguity. This revision modifies traditional values of the EM part of
the baryon mass splittings.

It is clearly valuable to independently determine either the strong
or EM contribution to the proton-neutron mass difference. In principle
this is achievable with lattice QCD. At this time, however, most
lattice simulations for the octet baryon masses are performed with 2+1
quark flavours, that is, with mass-degenerate light quarks: $m_u=m_d$.
Our analysis uses isospin-averaged lattice simulation
results\cite{Aoki:2008sm,Bietenholz:2011qq} to constrain chiral
perturbation theory expressions for the baryon masses. Because of the
symmetries of chiral perturbation theory, the only additional input
required to determine the strong contribution to the baryon mass
splittings is the up-down quark mass ratio $m_u/m_d$. The remainder of
this section is devoted to an illustration of this method.

\begin{figure}[!htbpf]
\begin{center}
\includegraphics[width=0.65\textwidth]{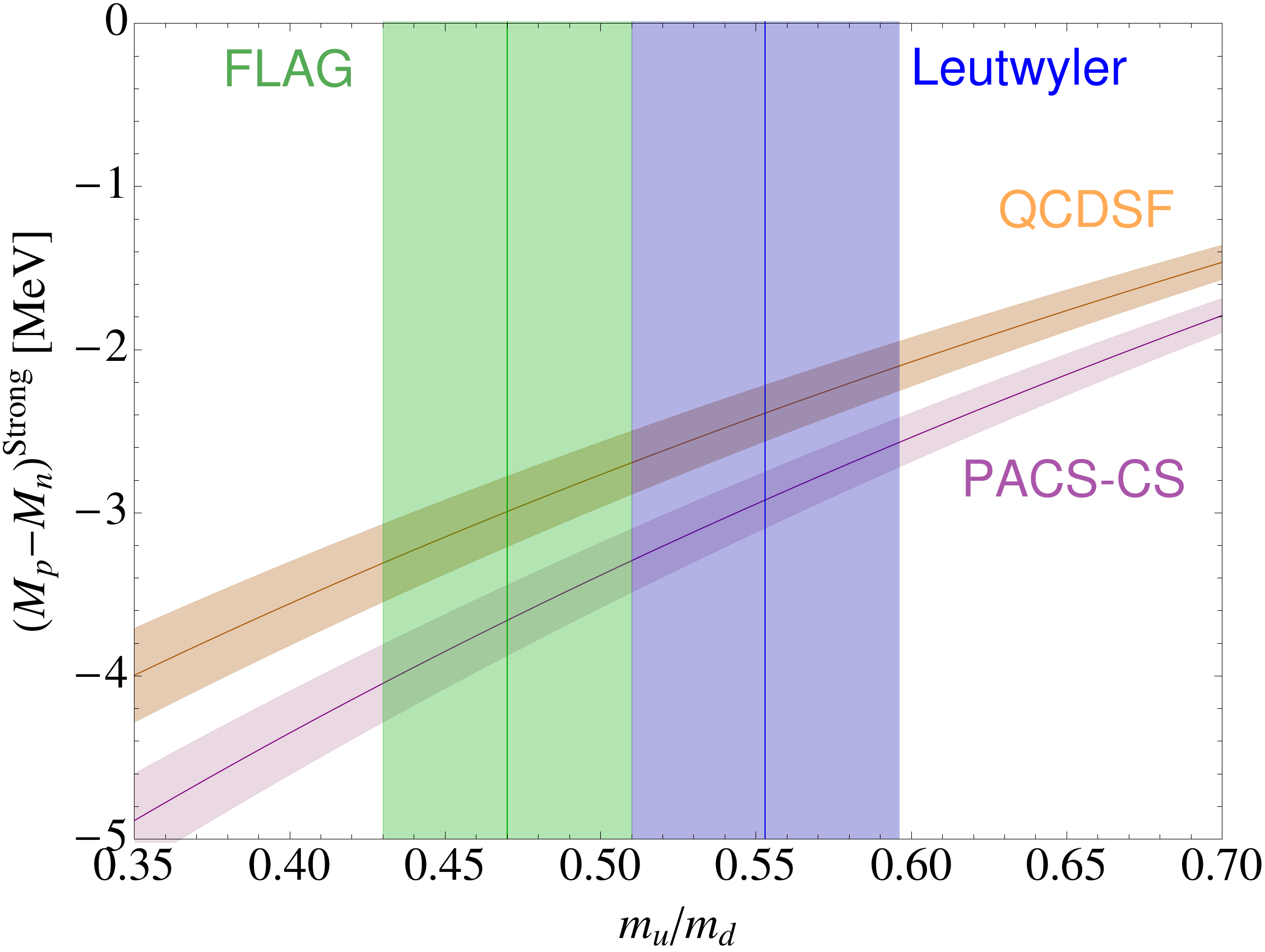}
\caption{Strong nucleon mass splitting from our analysis of two
  independent lattice simulations (QCDSF\cite{Bietenholz:2011qq} and
  PACS-CS\cite{Aoki:2008sm}), plotted against the quark mass ratio
  $m_u/m_d$. Phenomenological (Leutwyler\cite{Leutwyler:1996qg}) and
  lattice (FLAG\cite{Colangelo:2010et}) values for this ratio are
  shown.}
\label{fig:StrongQ}
\end{center}
\end{figure}

\begin{figure}[!htbpf]
\begin{center}
\includegraphics[width=0.65\textwidth]{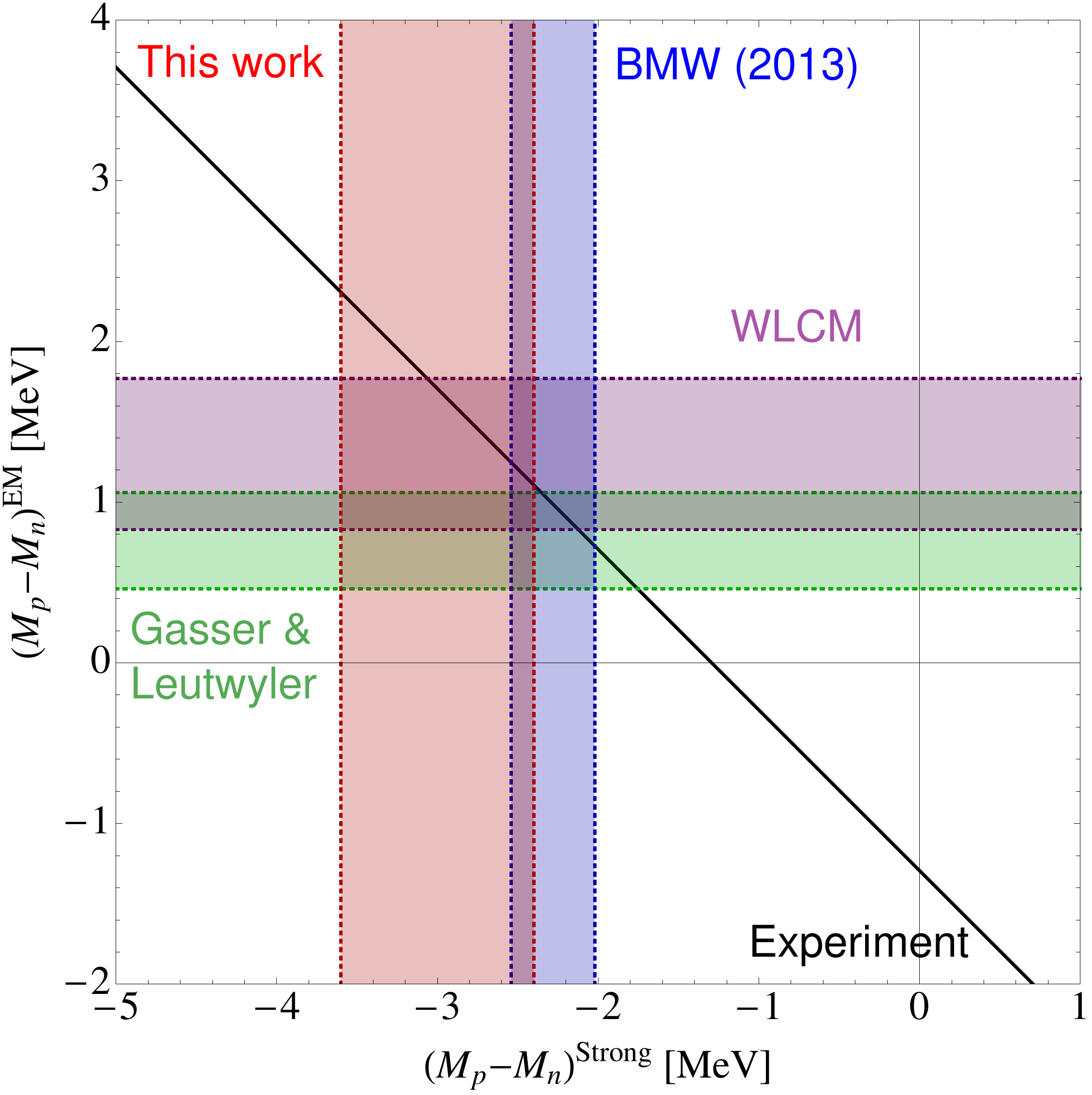}
\caption{Status of the nucleon mass splitting
  decomposition. Gasser-Leutwyler\cite{Gasser:1982ap} and
  WLCM\cite{WalkerLoud:2012bg} calculations of the electromagnetic
  contribution are compared with the strong contribution determined in
  this work\cite{Shanahan:2012wa} and by the BMW lattice
  collaboration\cite{Borsanyi:2013lga}. The black line indicates the
  experimental determination of the total mass
  difference\cite{Beringer:1900zz}.}
\label{fig:StrongEM}
\end{center}
\end{figure}

The usual meson-baryon Lagrangian can be written
\begin{align*} \nonumber
\mathcal{L}^B=&i\Tr \overline{\bf B}(v\cdot \mathcal{D}){\bf B} +2D \Tr \overline{\bf B}S^\mu \{A_\mu, {\bf B}\}+2F \Tr \overline{\bf B}S^\mu \left[ A_\mu, {\bf B} \right] \\ \nonumber
& {+ 2b_D\Tr \overline{\bf B}\{\mathcal{M}_q, {\bf B}\} +2b_F\Tr \overline{\bf B}\left[ \mathcal{M}_q, {\bf B} \right] }\\
& {+ 2\sigma_0 \Tr  \mathcal{M}_q\,\Tr \overline{\bf B}{\bf B}}.
\end{align*}
The $D$ and $F$ terms denote the meson--baryon interactions and
generate the nonanalytic quark mass dependence associated with quantum
fluctuations of the pseudo-Goldstone modes.
The explicit quark mass dependence is carried by the mass matrix
$\mathcal{M}_q$, which is related to only three undetermined
low-energy constants: $b_D$, $b_F$ and $\sigma_0$ (at this order).
With these constants determined by a fit to isospin-averaged
(2+1-flavour) lattice data, there are no new parameters in the
effective field theory relevant to CSV.
Combined with appropriate treatment of the CSV loop
corrections, our analysis of two independent
lattice simulations yields the charge symmetry-breaking derivative\cite{Shanahan:2012wa}
\begin{align*}
m_\pi^2\frac{d}{d\omega}(M_n-M_p)&=(20.3\pm 1.2)\mev\quad \textrm{[PACS-CS]}\\
m_\pi^2\frac{d}{d\omega}(M_n-M_p)&=(16.6\pm 1.2)\mev\quad \textrm{[QCDSF]}.
\end{align*}
Here the quark mass splitting is denoted by $\omega$, which is related to the quark mass ratio ($R=m_u/m_d$) by
\begin{equation}
\omega=\frac{1}{2}\frac{(1-R)}{(1+R)}m_{\pi\mathrm{(phys)}}^2.
\end{equation}
The dependence of our determination of $(M_p-M_n)^\text{Strong}$ on
the input quark mass ratio is indicated in Fig.~\ref{fig:StrongQ}.
In Figure \ref{fig:StrongEM} this analysis, where we consider both PACS-CS and Leutwyler results and allow for both Leutwlyer and FLAG values of the ratio $m_u/m_d$, is compared against a
recent strong mass splitting calculation of the BMW
Collaboration\cite{Borsanyi:2013lga} and the phenomenological
estimates of the electromagnetic self
energy\cite{Gasser:1982ap,WalkerLoud:2012bg}. Only for the purpose of
simplifying the graphic have we not shown other recent lattice QCD
estimates of the strong contribution to the mass splitting~\cite{Horsley:2012fw,deDivitiis:2011eh,Blum:2010ym,Beane:2006fk}.

\section{CSV parton distribution moments}
\label{sec:CSVParton}

The spin-independent CSV Mellin moments are defined as
\begin{align*}
\delta u^{m\pm} &= \int_0^1 dx x^m (u^{p\pm}(x)-d^{n\pm}(x))  \\
& = \langle x^m \rangle_u^{p\pm} - \langle x^m \rangle_d^{n\pm}, \\[8pt]
\delta d^{m\pm} &= \int_0^1 dx x^m (d^{p\pm}(x) - u^{n\pm}(x))  \\
&= \langle x^m \rangle_d^{p\pm}- \langle x^m \rangle_u^{n\pm},
\end{align*}
with similar expressions for the analogous spin-dependent terms $\delta\Delta q^\pm$. Here, the plus (minus) superscripts indicate C-even (C-odd) distributions $q^{\pm}(x)=q(x)\pm \overline{q}(x)$.

\begin{figure}[!ht]
\begin{center}
\includegraphics[width=0.68\textwidth]{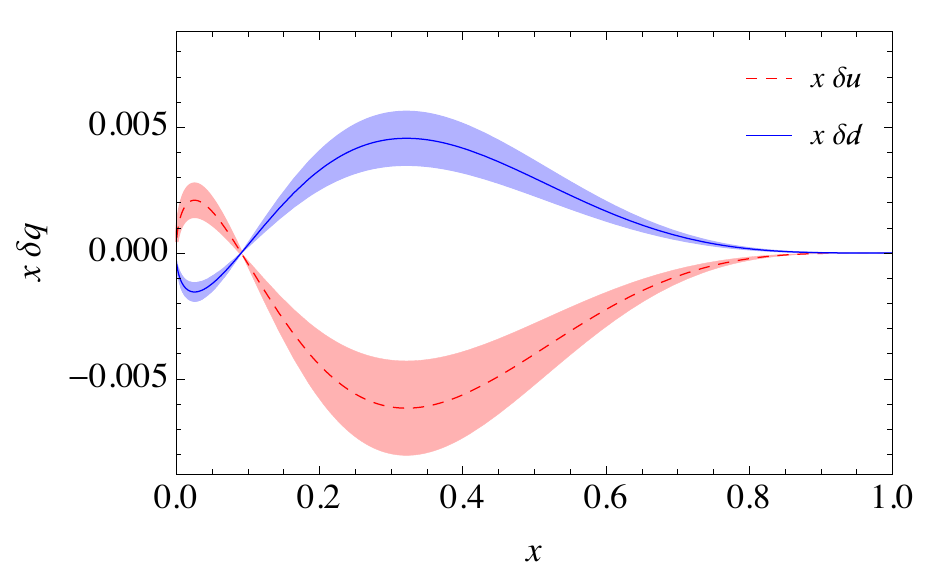}
\caption{Charge symmetry violating momentum fraction using simple phenomenological parameterisation $\delta q(x)=\kappa x^{-1/2}(1-x)^4(x-1/11)$ with normalisation determined from the lattice moment\cite{Shanahan:2013vla}.
\label{fig:csvpdf}}
\end{center}
\end{figure}

The first two spin-dependent and first spin-independent
lattice-accessible moments have recently been determined from
$2+1-$flavor lattice QCD by the QCDSF/UKQCD
Collaboration\cite{Cloet:2012db,Horsley:2010th}.
These original papers made first estimates for the amount of CSV in the parton moments by considering the leading flavour expansion about the SU(3) symmetric point\cite{Cloet:2012db,Horsley:2010th}.
In Ref.\cite{Shanahan:2013vla} we applied an SU(3) chiral expansion in the same fashion as the baryon mass expansion described above. This enabled us to extrapolate the results away from the SU(3) symmetric point to determine the CSV contribution at the physical quark masses.
Although this work only determines the lowest nontrival spin-independent
moment, we can infer the CSV distribution as
shown in Fig.~\ref{fig:csvpdf} by using the same parameterisation of the $x$ dependence
as Ref.~\cite{Martin:2003sk}.

This magnitude of charge symmetry breaking is found to be in agreement
with phenomenological MIT bag model
estimates\cite{Rodionov:1994cg,Londergan:2003ij}. This result is of particular
significance in the context of a new program to measure the (PVDIS)
asymmetry to high precision at Jefferson
Laboratory\cite{JLAB,Wang:2013kkc}. Further, the sign and magnitude of these results
suggest a 1-$\sigma$ reduction of the NuTeV
anomaly\cite{Bentz:2009yy}.

\section{Octet baryon spin fractions}
\label{sec:SpinFrac}
\begin{figure}
\label{fig:SpinFraction}
\begin{center}
\includegraphics[width=0.68\textwidth]{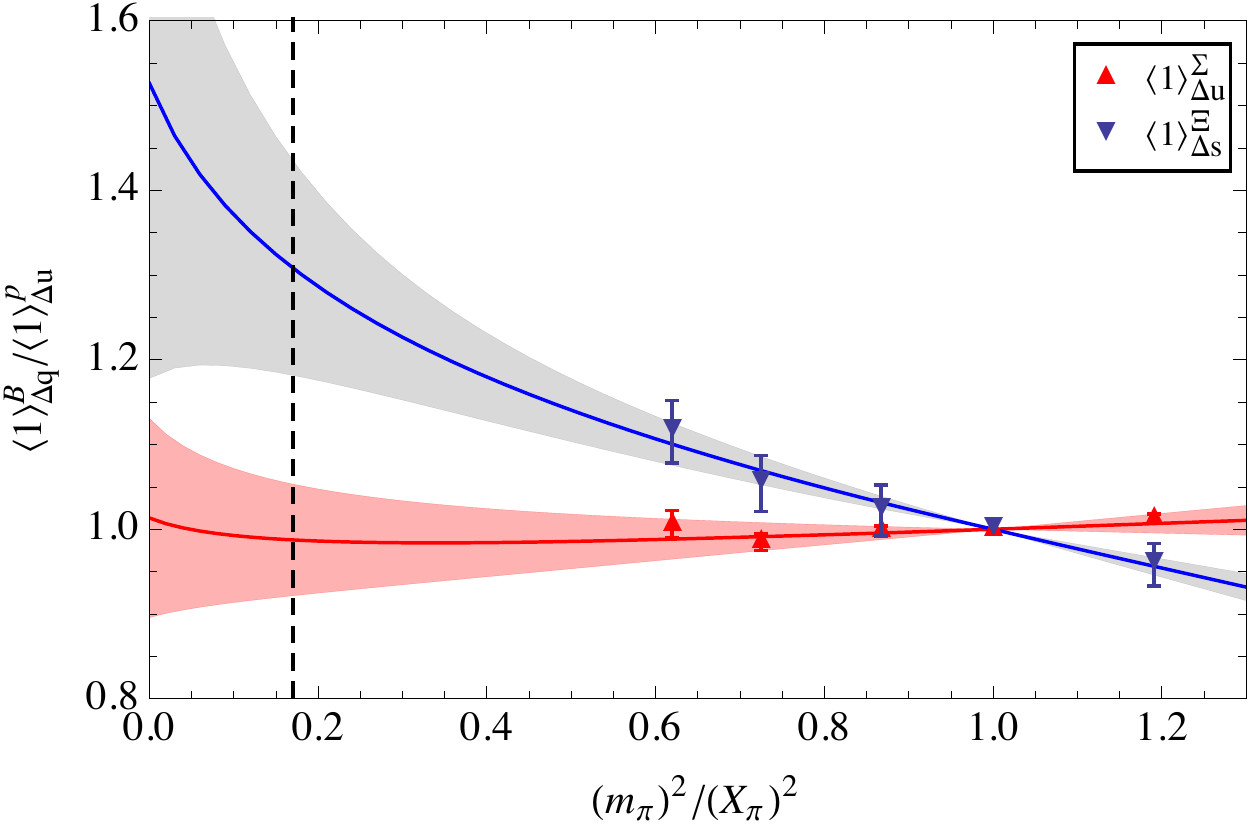}
\end{center}
\caption{Ratio of doubly-represented quark spin fractions in the octet baryons, taken from Ref.\cite{Shanahan:2013apa} . $X_\pi$ is the singlet quark mass.}
\end{figure}

In addition to using the chiral extrapolation of the previous section
to extract CSV effects, we have also determined the relative quark
spin fractions in the octet baryons\cite{Shanahan:2013apa}.
Figure~\ref{fig:SpinFraction}, taken from Ref.\cite{Shanahan:2013apa},
illustrates that the quark spin fraction is environment dependent. The
figure clearly highlights that this result is evident in the bare lattice
results, with considerable enhancement seen in the extrapolation to
the physical point.
Clearly, any candidate explanation of the proton spin problem must
allow for the fraction of spin carried by the quarks to be dependent
on baryon structure.

This finding is supported by a Cloudy Bag Model calculation, which
includes relativistic and one-gluon-exchange
corrections~\cite{Myhrer:1988ap,Myhrer:2007cf,Schreiber:1988uw}.
Within this model, the observed variation in quark spin arises from
the meson cloud correction being considerably smaller in the $\Xi$
than in the nucleon. That, combined with the less relativistic motion
of the heavier strange quark, results in the total spin fraction in
the $\Xi$ being significantly larger than in the nucleon.

\section{Conclusion}

The effects of charge symmetry violation (CSV) are becoming
increasingly significant in precision studies of the Standard
Model. Recent results, based on $2+1-$flavor lattice QCD simulations,
unambiguously resolve CSV in the quark Mellin moments. These results
reduce the NuTeV anomaly from $3\sigma$ to $2\sigma$ and could improve
the sensitivity of Standard Model tests such as the PVDIS program at
Jefferson Laboratory. The same lattice QCD studies show that the
fraction of baryon spin carried by the quarks is structure-dependent,
rather than universal across the baryon octet.

\section{Acknowledgements}
This work was supported by the University of Adelaide and the
Australian Research Council through through the ARC Centre of
Excellence for Particle Physics at the Terascale and grants FL0992247
(AWT), DP110101265 (RDY) and FT120100821 (RDY).

\bibliographystyle{ws-procs9x6}

\end{document}